\documentclass[english,aps,manuscript]{revtex4}
\usepackage[T1]{fontenc}
\usepackage[latin9]{inputenc}
\usepackage{graphicx}
\usepackage{amssymb}

\makeatletter

\providecommand{\tabularnewline}{\\}

\usepackage{babel}
\makeatother

\begin{document}

\title{Single Production of Fourth Family Quarks at the LHC}

\author{O. \c{C}ak\i{}r}

\email{ocakir@science.ankara.edu.tr}

\affiliation{Ankara University, Faculty of Sciences, Department of Physics, 06100, Tandogan, Ankara,
Turkey}

\author{\.{I}. T. \c{C}ak\i{}r}

\email{iturk@science.ankara.edu.tr}

\affiliation{Ankara University, Faculty of Sciences, Department of Physics,
Tandogan, Ankara, Turkey}

\author{H. Duran Y\i{}ld\i{}z}

\email{hyildiz@dumlupinar.edu.tr}

\affiliation{Dumlupinar University, Faculty of Art and Sciences, Department of
Physics, Central Campus - K\"{u}tahya, Turkey }

\author{R. Mehdiyev}

\email{rmehdi@fnal.gov}

\affiliation{University of Texas at Austin, Department of Physics, Austin, Texas,
Texas 78712-0264, USA}

\address{on leave of absence from Institute of Physics, Azerbaijan National
Academy of Sciences, 370143, Baku, Azerbaijan}

\begin{abstract}
We study the single production of fourth family quarks through the
process $pp\to Q'jX$ at the Large Hadron Collider (LHC). We have
calculated the decay widths and branching ratios of the fourth family
quarks ($b'$ and $t'$) in the mass range 300-800 GeV. The cross
sections for the signal and background processes have been calculated
in a Monte Carlo framework. It is shown that the LHC can discover
single $t'$ and $b'$ quarks if the CKM matrix elements $|V_{t'q}|,|V_{qb'}|\gtrsim0.01$. 
\end{abstract}


\maketitle

\section{Introduction}

The Standard Model (SM) of the electroweak and color interactions
does not predict the number of the fermion families. However, due
to the limits coming from strong coupling in QCD, the number of quark
flavors should be less than eighteen. The electroweak precision measurements
done by LEP experiments imply that the number of the light neutrinos
is equal to three \cite{yao06}. The replications of three quark
and lepton families remain mystery as part of the flavor problem.
A possible fourth family can play a crucial role to understand the
flavor structure of the standard theory. The experiments at Tevatron
have already constrained the masses of fourth family quarks. The CDF
has the strongest bound on the mass of $t'$ quark as $m_{t'}>258$
GeV at $95\%$ CL \cite{cdf05}. This analysis applies to $t'$ pair
production independent of the CKM matrix elements $V_{t'q}$. In particular,
CDF obtains a bound on the mass of $b'$ quark $m_{b'}>268$ GeV \cite{aaltonen07}
taking $BR(b'\to bZ)=1$. Recent reexamination of the bounds on the
masses of the fourth family quarks assumes $m_{Q'}>255$ GeV \cite{hung08}.
There exist some parameter space (mass-mixing angle) of the fourth
family quarks which could be explored at future collider searches.
The upcoming experiments at the Large Hadron Collider (LHC) are able
to probe heavy quarks up to a heavier mass range accessible to future
experiments. Recently, the pair production signals have been considered
in Ref. \cite{holdom06} with a similar mass range what we consider
here. Because of the uncertainties on the CKM matrix elements, there
still remains an open door for the fourth family quarks and their
mixings with the other three families. Recent electroweak precision
S and T parameter analysis also supports to idea of the existence
of the fourth family quarks \cite{kribs07}.

A prediction for the fourth family fermion masses as well as the CKM
mixings has been presented in \cite{celikel95,atag96}. Pair production
of the fourth SM family quarks has been studied at the LHC \cite{arik98,atlas99}.
Recently, an update has been made on the analysis of the signal and
background for the pair production of fourth family quarks within
the ATLAS detector \cite{gokhan07}.

In this paper, we study the single production (less studied compared
to pair production) of the fourth family quarks through the process
$pp\to b'(\bar{b'})jX$ and $pp\to t'(\bar{t'})jX$ at the LHC. We
have calculated the cross sections of signals and corresponding backgrounds
in the Monte Carlo (MC) framework. The decay widths and branching
ratios of the fourth family quarks ($b'$ and $t'$) are calculated
in the mass range 300-800 GeV. We analyze the potential of the LHC
for probing the fourth family quarks providing three different parametrizations
of the CKM matrix elements $|V_{t'q}|,|V_{qb'}|$.

\section{Interaction Lagrangian with Fourth Family Quarks }

In the standard model of the electroweak interactions, quarks can
couple to charged weak currents by the exchange of $W^{\pm}$ boson
and neutral weak currents by $Z^{0}$ boson. The left-handed (right-handed)
quark fields transform as doublets (singlets) under the group $SU(2)$.
We consider an enlarged framework of the SM to include fourth family
quarks. The interaction of the fourth family quarks $Q_{i}^{'}(t',b')$
and the quarks $q_{i}$ via the SM gauge bosons ($\gamma,g,Z^{0},W^{\pm}$)
is given by

\begin{eqnarray}
L & = & -g_{e}\sum_{Q_{i}'=b',t'}Q_{ei}\overline{Q}_{i}'\gamma^{\mu}Q_{i}'A_{\mu}\nonumber \\
 &  & -g_{s}\sum_{Q_{i}'=b',t'}\overline{Q}_{i}'T^{a}\gamma^{\mu}Q_{i}'G_{\mu}^{a}\nonumber \\
 &  & -\frac{g}{2cos\theta_{W}}\sum_{Q_{i}'=b',t'}\overline{Q}_{i}'\gamma^{\mu}(g_{V}^{i}-g_{A}^{i}\gamma^{5})Q_{i}'Z_{\mu}^{0}\nonumber \\
 &  & -\frac{g}{2\sqrt{2}}\sum_{Q_{j\neq i}'=b',t'}V_{ij}\overline{Q}_{i}'\gamma^{\mu}(1-\gamma^{5})q_{j}W_{\mu}^{\pm}\label{eq:1}\end{eqnarray}
where $g_{e},g$ are electro-weak coupling constants, and $g_{s}$
is the QCD coupling constant. $A_{\mu},G_{\mu},Z_{\mu}^{0}$ and $W_{\mu}^{\pm}$
are the vector fields for photon, gluon, $Z^{0}$-boson and $W^{\pm}$-boson,
respectively. $Q_{ei}$ is the electric charge of fourth family quarks;
$T^{a}$ are the Gellman matrices. $g_{V}$ and $g_{A}$ are the vector
and axial-vector type couplings of the neutral weak current. Finally,
$V=Y^{u\dagger}Y^{d}$(where $Y^{u,d}$ being the Yukawa couplings)
is the corresponding $4\times4$ CKM matrix:

\begin{equation}
V=\left(\begin{array}{cccc}
V_{ud} & V_{us} & V_{ub} & V_{ub'}\\
V_{cd} & V_{cs} & V_{cb} & V_{cb'}\\
V_{td} & V_{ts} & V_{tb} & V_{tb'}\\
V_{t'd} & V_{t's} & V_{t'b} & V_{t'b'}\end{array}\right)\label{eq:2}\end{equation}
The constraint on the mixing between the quarks comes from the unitarity
of the CKM matrix in the SM. These constraints appear to be well satisfied
experimentally for the three-family case and the elements of the $3\times3$
sub-matrix are well tested by various processes \cite{yao06}. The
elements in the fourth row and column are constrained by flavor physics,
and the flavor changing neutral current (FCNC) effects are suppressed
by Glashow-Iliopoulos-Maiani (GIM) mechanism. A $4\times4$ quark
mixing matrix $V$ satisfy the unitarity conditions 

\begin{equation}
\sum_{\alpha=1}^{4}|V_{i\alpha}|^{2}=\sum_{j=1}^{4}|V_{j\beta}|^{2}=1\label{eq:3}\end{equation}
for $i,j=u,c,t,t'$ and $\alpha,\beta=d,s,b,b'$. In addition to these
constraints, if the following constraints are also satisfied 

\begin{equation}
\sum_{\alpha=1}^{4}V_{i\alpha}V_{j\alpha}^{*}=\sum_{j=1}^{4}V_{j\alpha}V_{j\beta}^{*}=0\label{eq:4}\end{equation}
for all indices ($i\neq j$ in the first sum and $\alpha\neq\beta$
in the second sum) the triangles will have the same areas. First,
second and third rows of the matrix $V$ with the measurement from
currently available experiments can be calculated as $|V_{ub'}|^{2}=0.0008\pm0.001$1,
$|V_{cb'}|^{2}=0.0295\pm0.1780$ and $|V_{tb'}|^{2}=0.4054\pm0.3696$.
For the first three columns we calculate $|V_{t'd}|^{2}=-0.001\pm0.005$,
$|V_{t's}|^{2}=0.0315\pm0.1780$ and $|V_{t'b}|^{2}=0.4053\pm0.3696$.
The mixing between the third and fourth family is less constrained.
A lower limit can be obtained from the single production of top quarks,
such as $|V_{tb}|>0.68$ at $95\%$ C.L. \cite{abazov07}. If there
is a mass degeneracy between $b'$ and $t'$ quarks, the two body
decay occurs mostly into third generation quarks via charged currents.

\section{Production Cross Section}

\begin{figure}
\includegraphics{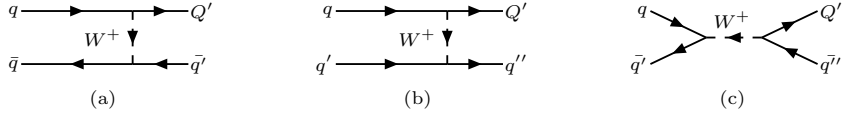}

\caption{The diagrams for single production of fourth family quarks. \label{fig1} }

\end{figure}

The fourth family quarks $b'$ and $t'$ can be produced singly at
the LHC. There are relevant processes $q\bar{q}\to Q'\bar{q'}$, $qq'\to Q'q''$
and $q\bar{q'}\to Q'\bar{q''}$ for the single production as shown
in Fig. \ref{fig1}. As can be seen from Eq. (1) and Fig. \ref{fig1}
the single production cross section strongly depends on the CKM elements
$V_{qb'}$ or $V_{t'q}$. The cross sections for the subprocesses
in Fig. \ref{fig1} are summed to make $"Q'+jet"$ analysis in the
final state. The total cross section for the process $pp\to Q'jX$
is expressed by 

\begin{figure}
\includegraphics{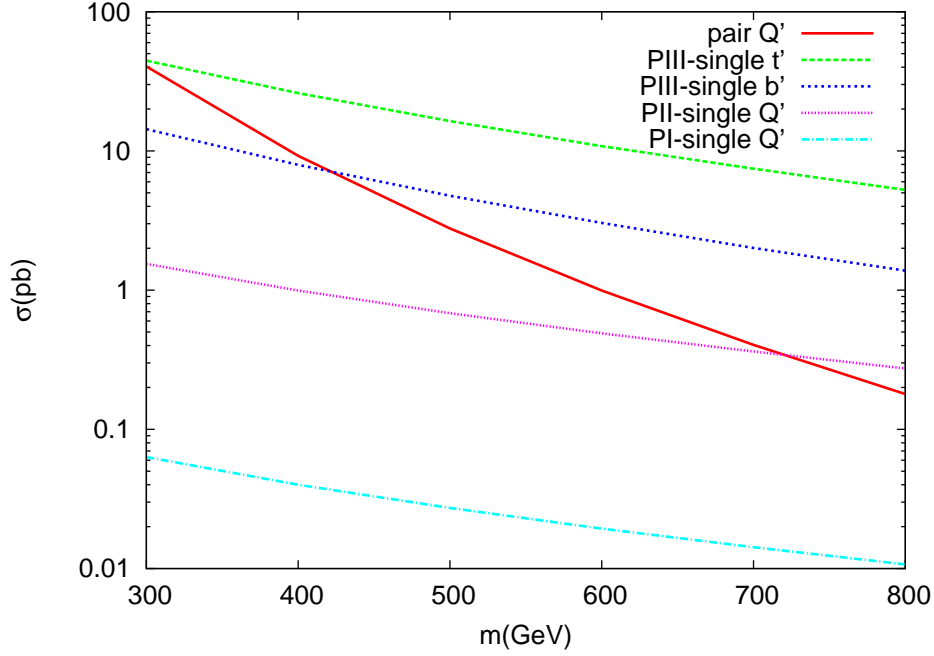}

\caption{The cross sections for the single and pair production of $b'$ and
$t'$ quarks at LHC for different parametrizations of the CKM matrix
elements $V_{t'q},V_{qb'}$. \label{fig2}}

\end{figure}

\begin{figure}
\includegraphics{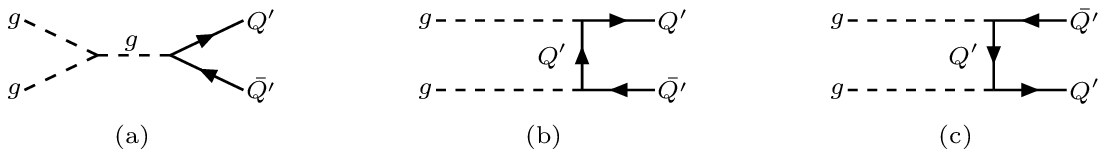} \includegraphics{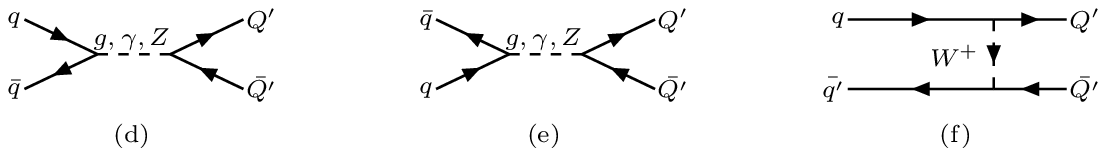}

\caption{The diagrams for the pair production of fourth family quarks through
gluon-gluon fusion and quark anti-quark annihilation processes. \label{fig3}}

\end{figure}

\begin{equation}
\sigma=\sum_{i,j}\int_{y_{min}}^{1}dy\int_{y}^{1}\frac{dx}{x}f_{q_{i}/p}(x,Q^{2})f_{q_{j}/p}(y/x,Q^{2})\hat{\sigma}(ys)\label{eq:5}\end{equation}
where the lower limit of the first integral is chosen as $y_{min}=m_{Q'}^{2}/s$.
The parton distribution function (PDF) $f_{q/p}(x,Q^{2})$ is taken
from the CTEQ6M \cite{cteq6m} with $Q^{2}=m_{Q'}^{2}$ for the signal
and $Q^{2}=\hat{s}$ for the background. In order to compare the magnitude
of single production cross section with the pair production we plot
Fig. \ref{fig2} depending on the mass of fourth family quarks. The
pair production of fourth family quarks at the LHC is mainly due to
gluon-gluon fusion and quark-antiquark annihilation processes \cite{gokhan07}.
In addition, there is also a contribution from the $W$-boson exchange
in the $t$-channel when the CKM matrix elements $V_{t'q},V_{qb'}$
are taken large, in which the relevant diagrams are shown in Fig.
\ref{fig3}. We use three parametrizations called PI, PII and PIII
for the CKM elements corresponding to $V_{Q'i}=V_{iQ'}=0.01$, $V_{Q'i}=V_{iQ'}=0.05$
and $V_{ub'}=0.044$, $V_{cb'}=0.46$, $V_{t'd}=0.063$, $V_{t's}=0.46$,
$V_{t'b}=0.47$, respectively. The reason for choosing these options
is that only the upper limits on the CKM elements could be well predicted
from the current experimental data, and they can be relaxed as $1\sigma$
over the average values. For a parametrization with equal strength
$V_{iQ'}=V_{Q'i}$, single production cross section becomes comparable
to the pair production cross section for the value of CKM matrix elements
$V_{t'q},V_{qb'}=0.25-0.04$ in the mass range $m_{Q'}=300-800$ GeV,
as shown in Fig. \ref{fig2}. The cross section for single $t'$ and
$b'$ production is almost the same for the parametrization PI and
PII in the interested mass range.

The decay channels of fourth family quarks are strongly related to
the non-diagonal matrix elements $V_{t'q},V_{qb'}$ and the diagonal
$V_{t'b'}$. The production cross section depending on the mass of
fourth family quarks is given in Table \ref{table1} for the parametrizations
PI-PIII. These values are obtained by using CalcHEP program \cite{pukhov99}.
In the first and second parametrizations the cross-sections are low
as expected from the fact that single production cross section simply
scales with $|V_{t'q}|^{2}$ and $|V_{qb'}|^{2}$. All the related
vertices are implemented into the program and the model files can
be found at \cite{ocakir07}. In order to compare the range of the
error in cross sections coming from the source of parton distributions
CTEQ6L and CTEQ6M at $Q^{2}=m_{Q'}^{2}$, we present the total cross
section in Table \ref{table2} for single and pair production of fourth
family quarks at the LHC. As can be seen from this table the relative
difference in pair production is higher than in the single production
mechanism. The single production cross sections differ by about $10\%$
at $m_{Q'}=300$ GeV for two options of CTEQ distribution, while pair
production differs at most $\sim18\%$. For higher mass values the
difference between the two choices gets smaller. The ratios $\sigma(pp\to t'jX)/\sigma(pp\to tjX)$
and $\sigma(pp\to t'\bar{t}'X)/\sigma(pp\to t\bar{t}X)$ can also
be used to reduce the uncertainties from the PDFs. The cross sections
for the single and pair production of top quarks at the LHC are: $140.94$
pb for single production and $492.36$ pb for pair production when
the CTEQ6L is used at the scale $Q^{2}=m_{t}^{2}$. 

\begin{table}
\caption{The cross section in pb for the single production of fourth family
quarks $b'(\bar{b'})$ and $t'(\bar{t'})$ at the LHC. The parametrization
PI-PIII are explained in the text. The numbers in the paranthesis
denote the antiparticle case. Here, we have used CTEQ6M at the scale
$Q^{2}=m_{Q'}^{2}$. \label{table1}}

\begin{tabular}{|c|c|c|c|c|c|c|}
\hline 
Process$\to$ & \multicolumn{3}{c|}{$pp\to b'(\bar{b'})jX$} & \multicolumn{3}{c|}{$pp\to t'(\bar{t'})jX$}\tabularnewline
\hline 
Mass(GeV) & PI & PII & PIII & PI & PII & PIII\tabularnewline
\hline 
300 & 0.0617(0.0250) & 1.54(0.626) & 14.34(20.31) & 0.0631(0.0234) & 1.579(0.585) & 44.56(25.70)\tabularnewline
\hline 
400 & 0.0398(0.0149) & 0.995(0.372) & 7.95(11.77) & 0.0401(0.0134) & 1.003(0.335) & 26.06(14.44)\tabularnewline
\hline 
500 & 0.0273(0.0094) & 0.683(0.236) & 4.77(7.31) & 0.0272(0.0082) & 0.681(0.206) & 16.36(8.75)\tabularnewline
\hline 
600 & 0.0196(0.0063) & 0.489(0.157) & 3.03(4.78) & 0.0194(0.0053) & 0.484(0.133) & 10.82(5.59)\tabularnewline
\hline 
700 & 0.0145(0.0043) & 0.361(0.109) & 2.01(3.24) & 0.0142(0.0036) & 0.356(0.0897) & 7.42(3.71)\tabularnewline
\hline
800 & 0.0109(0.0031) & 0.274(0.0772) & 1.38(2.26) & 0.0107(0.0025) & 0.268(0.0621) & 5.25(2.54)\tabularnewline
\hline
\end{tabular}
\end{table}

\begin{table}
\caption{The total cross section in pb for the pair production of fourth family
quarks $Q'$ at the LHC. The values are given for two options of the
parton distribution function CTEQ6 with $Q^{2}=m_{Q'}^{2}$.\label{table2}}

\begin{tabular}{|c|c|c|c|c|}
\hline 
$Q^{2}=m_{Q'}^{2}$ & \multicolumn{2}{c|}{$pp\to Q'jX$(PI)} & \multicolumn{2}{c|}{$pp\to Q'\bar{Q}'X$(PI)}\tabularnewline
\hline 
Mass(GeV) & CTEQ6L & CTEQ6M & CTEQ6L & CTEQ6M\tabularnewline
\hline 
300 & 0.056 & 0.062 & 34.34 & 40.36\tabularnewline
\hline 
400 & 0.036 & 0.040 & 7.63 & 9.21\tabularnewline
\hline 
500 & 0.025 & 0.027 & 2.27 & 2.77\tabularnewline
\hline 
600 & 0.018 & 0.019 & 0.82 & 0.99\tabularnewline
\hline 
700 & 0.014 & 0.014 & 0.33 & 0.40\tabularnewline
\hline
800 & 0.011 & 0.011 & 0.15 & 0.18\tabularnewline
\hline
\end{tabular}
\end{table}

\section{Decay Widths and Branchings}

If $m_{t'}>m_{b'}$, then $t'$ and $b'$ quarks can decay as

\begin{equation}
\begin{array}{ccccccccccc}
t' & \to & W^{+}b'\\
 & \to & W^{+}b &  &  &  &  &  & b' & \to & W^{-}t\\
 & \to & W^{+}q(d,s) &  &  &  &  &  &  & \to & W^{-}q(u,c)\end{array}\label{eq:6}\end{equation}
If $m_{b'}>m_{t'}$, the $b'$ and $t'$ quarks can decay as

\begin{equation}
\begin{array}{ccccccccccc}
b' & \to & W^{-}t'\\
 & \to & W^{-}t &  &  &  &  &  & t' & \to & W^{+}b\\
 & \to & W^{-}q(u,c) &  &  &  &  &  &  & \to & W^{+}q(d,s)\end{array}\label{eq:7}\end{equation}
In the first lines of the equations (\ref{eq:6}) and (\ref{eq:7}),
two body decay $Q_{i}'\to Q_{j}'+W^{\pm}$ takes place with a virtual
(real) $W$ boson depending on the mass difference $|m_{t'}-m_{b'}|\lesssim80$
($>80$) GeV. In the analysis, both two body decays and three body
decays (via virtual $W^{\pm*}$ boson) are considered. Fourth family
quarks will decay into third family quarks associated with a $W^{\pm}$
boson as shown in the second line of the Eq. (\ref{eq:6}) and (\ref{eq:7}).
Last lines of the above equations denote the fourth family quark decays
into light jets. 

The branching ratios for the $b'$ decays according to the parametrization
PIII are given in Table \ref{table3}. As can be seen from this table,
the fractions are dominant for $b'\to W^{-}c$ and $b'\to W^{-}t$
channels. For the other parametrization, the branchings change slightly
in the interested mass range, for example: BR$(b'\to W^{-}u(c))=36.6(33.4)\%$
and BR$(b'\to W^{-}t)=26.9(33.3)\%$ at $m_{b'}=300(800)$ GeV. Concerning
the options PI and PII, the $t'$ branchings BR$(t'\to W^{+}q)=33.3\%$
remain the same for all channels in the considered mass range, while
they are $51\%(W^{+}b)$, $48\%(W^{+}s)$ and $1\%(W^{+}d)$ for the
option PIII. In Table \ref{table3} the $b'$ decay modes and fractions
are displayed. We have used these modes of fractions in order to calculate
statistical significances (SS) as explained in the next section. 

\begin{table}
\caption{The branching ratio ($\%$) for the $b'$ decays according to the
parametrization PIII. For the other parametrizations PI and PII the
branchings change very small in the considered mass range. \label{table3}}

\begin{tabular}{|c|c|c|c|}
\hline 
Mass(GeV) & $W^{-}u$ & $W^{-}c$ & $W^{-}t$\tabularnewline
\hline 
300 & 0.5 & 56.3 & 43.2\tabularnewline
\hline 
400 & 0.5 & 50.2 & 49.3\tabularnewline
\hline 
500 & 0.5 & 50.3 & 49.2\tabularnewline
\hline 
600 & 0.5 & 48.9 & 50.6\tabularnewline
\hline 
700 & 0.5 & 48.8 & 50.7\tabularnewline
\hline 
800 & 0.4 & 48.8 & 50.8\tabularnewline
\hline
\end{tabular}
\end{table}

\begin{figure}
\includegraphics{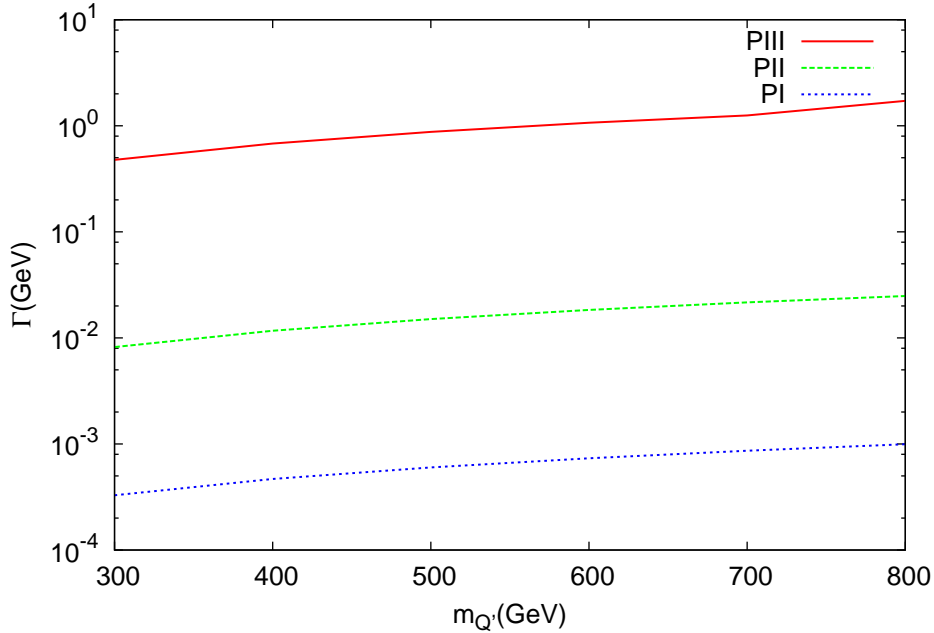}

\caption{Decay width of fourth family quarks for the parametrization PI-PIII.
\label{fig4}}

\end{figure}

In the case of mass degeneracy $m_{t'}=m_{b'}$ both $t'$ and $b'$
quarks decay into SM three family quarks with a charged current interaction.
The total decay widths depending on the mass are given in Fig. \ref{fig4}
for three parametrization PI-PIII of the CKM matrix elements. The
total widths are increasing smoothly with increasing masses. The option
PIII has the highest total widths among the other options PI and PII.
For example, at PIII with $m_{Q'}=700$ GeV the decay width $\Gamma_{Q'}=1.5$
GeV which is also comparable to that of the top quark.

\section{Signal and Background}

\begin{figure}
\includegraphics{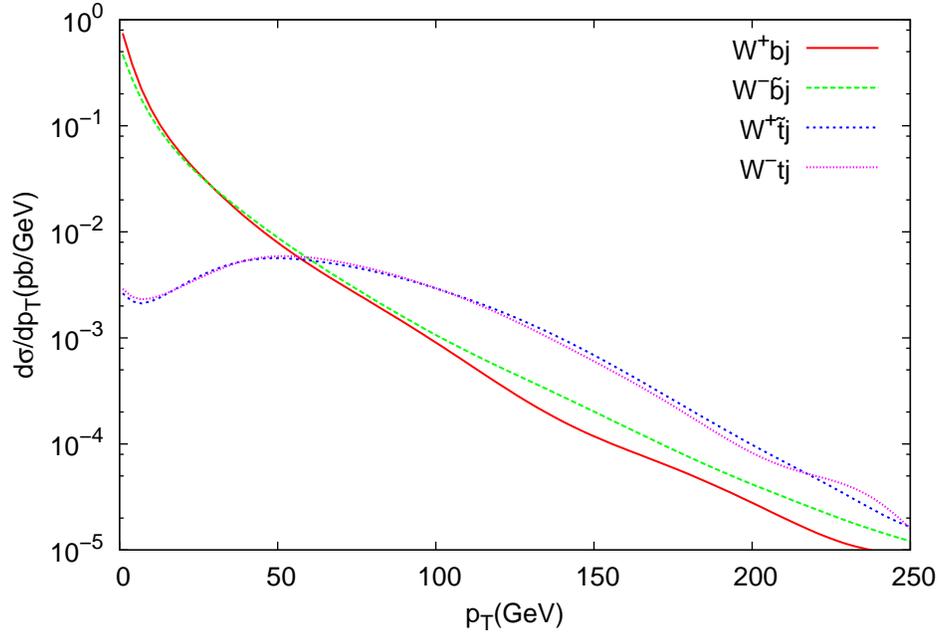}

\caption{The differential cross section depending on the transverse momentum
of final state jet. The solid line (red) and dashed line (green) correspond
to $W^{+}bj$ and $W^{-}\bar{b}j$ backgrounds, respectively; the
dotted and dot-dashed lines refer to the backgrounds $W^{+}\bar{t}j$
and $W^{-}tj$ , respectively. \label{fig5}}

\end{figure}

\begin{figure}
\includegraphics{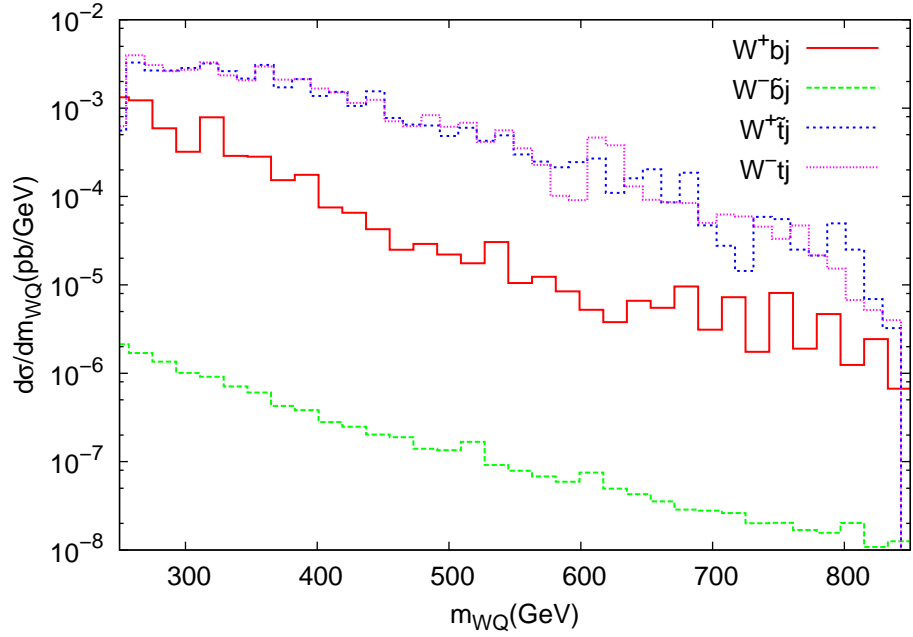}

\caption{Invariant mass distributions for $W^{\pm}Q$ final state, the solid
(red) and dashed (green) lines correspond to $W^{+}bj$ and $W^{-}\bar{b}j$
backgrounds, while dotted (blue) and dotted-dashed (purple) lines
correspond to $W^{+}\bar{t}j$ and $W^{-}tj$. \label{fig6}}

\end{figure}

\begin{figure}
\includegraphics{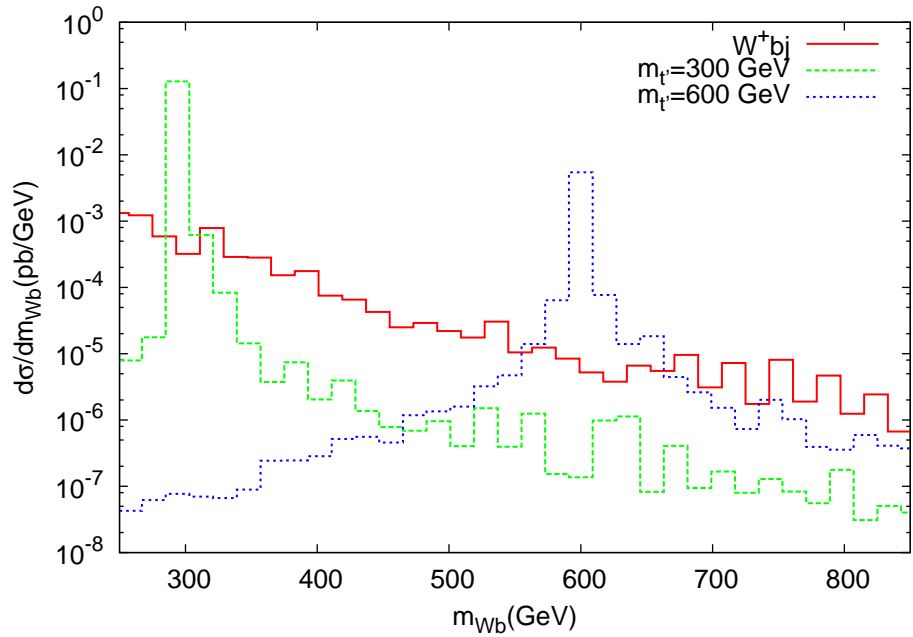}

\caption{The invariant mass distribution for $Wb$ signal from $t'$ decay
and the corresponding background. \label{fig7}}

\end{figure}

The signal which we consider here for $t'$ and $b'$ single production
include the decay chains $t'\to W^{+}b\to l^{+}\nu_{l}b$ (or $l^{+}\nu_{l}j$)
and $b'\to W^{-}t\to l^{-}\bar{\nu_{l}}l^{+}\nu_{l}b$ (or $l^{-}\bar{\nu_{l}}l^{+}\nu_{l}j$),
respectively. The anti-particle $\bar{t'}$ decays similarly but with
an opposite charged lepton in the final state. However, $b'$ or $\bar{b'}$
quarks decay eventually into two oppositely charged leptons, hence
they should be taken into account together to enhance the statistics.
Since a $b$-jet can be identified with a good efficiency, we consider
$b$-jets in the final state. In our calculations we take into account
leptonic decays of the $W$ boson ($W\to l\nu_{l}$, where $l=e,\mu$).
The corresponding backgrounds contain a charged lepton, b-jet and/or
other jets and missing energy for the single $t'$ production signal;
two oppositely charged leptons, b-jet and/or other jets and missing
energy for the single $b'$ production signal. We calculate the cross
sections for the $W^{\pm}Z$, $W^{+}W^{-}$, $ZZZ,$ $ZZW^{\pm},$
$W^{+}W^{-}Z$, $W^{+}W^{-}W^{\pm}$ and $Wbj$, $Wtj$ backgrounds.
These backgrounds are calculated by using CalcHEP with the PDF CTEQ6M
at $Q^{2}=\hat{s}$. The results are given in Table \ref{table4}
for the backgrounds including two or three weak bosons. These backgrounds
give relatively small cross section when they are multiplied by the
corresponding branching ratios for the interested channels. The backgrounds
including $Wbj$ and/or $Wtj$ in the final state need to be studied
in detail to see the signal over the background, while the others
give relatively smaller cross sections. The background process $pp\to WbjX$
includes the single production of top quarks as well. By applying
an invariant mass cut on the $m_{Wb}>200$ GeV this background can
be reduced. The transverse momentum distribution of the final jet
for the background processes $pp\to WQjX$ is shown in Fig. \ref{fig5}.
The $p_{T}$ distributions of the jet for $W^{+}bj$ and $W^{-}\bar{b}j$
backgrounds have similar shape, while it differs from the backgrounds
$W^{+}\bar{t}j$ and $W^{-}tj$. In order to reduce the relevant backgrounds
and preserve the signal, we applied the acceptance cuts $p_{T}^{j}>20$
GeV for the final state jets. The invariant mass distribution $m_{WQ}$
is given in Fig. \ref{fig6}. For the interested invariant mass range
the backgrounds $W^{+}\bar{t}j$ and $W^{-}tj$ contribute dominantly.
We also apply an invariant mass cut as $|m_{t'}-m_{W^{+}b}|<10-20$
GeV depending on the $t'$ mass, and similarly, $|m_{b'}-m_{W^{-}t}|<10-20$
GeV for the $b'$ signal. In this case, we obtain a significant reduction
on the cross section of the background as shown in Table \ref{table5}.
The reconstructed invariant mass spectrum, $m_{Wb}$, for the $t'$
signal (where $m_{t'}=300$ GeV and $600$ GeV) and the corresponding
background are given in Fig. \ref{fig7}. 

\begin{table}
\caption{The cross-section in pb for the background relevant to the single
$b'(\bar{b'})$ and $t'(\bar{t'})$ productions as discussed in the
text. The results are given for the backgrounds including two and
three weak bosons in the final state. \label{table4}}

\begin{tabular}{|c|c|}
\hline 
Background & $\sigma(pb)$\tabularnewline
\hline
$ZZZ$ & $1.111\times10^{-2}$\tabularnewline
\hline 
$ZZW^{+}$ & $2.050\times10^{-2}$\tabularnewline
\hline 
$ZZW^{-}$ & $1.091\times10^{-3}$\tabularnewline
\hline 
$W^{+}W^{-}W^{+}$ & $8.826\times10^{-2}$\tabularnewline
\hline
$W^{+}W^{-}W^{-}$ & $4.463\times10^{-2}$\tabularnewline
\hline
$W^{+}W^{-}Z$ & $1.033\times10^{-1}$\tabularnewline
\hline
$W^{+}Z$ & $1.868\times10^{1}$\tabularnewline
\hline
$W^{-}Z$ & $1.169\times10^{1}$\tabularnewline
\hline
$W^{+}W^{-}$ & $8.367\times10^{1}$\tabularnewline
\hline
\end{tabular}
\end{table}

\begin{table}
\caption{The cross section in pb for $pp\to W^{-}tjX(W^{+}\bar{t}jX)$ and
$pp\to W^{+}bjX(W^{-}\bar{b}jX)$ background with (w) and without
(o) cuts $p_{T}^{j}>20(50)$ GeV in the different mass range of $\Delta m_{Wt}$
or $\Delta m_{Wb}$ (numbers shown in the first line are in GeV) .
\label{table5}}

\begin{tabular}{|c|c|c|c|c|c|c|c|}
\hline 
Process & no cut & $290-310$ & $390-410$ & $485-515$ & $585-615$ & $680-720$ & $780-820$\tabularnewline
\hline 
$pp\to W^{+}\overline{t}jX$ & $9.771$ & $0.444(0.238)$ & $0.204(0.125)$ & $0.132(0.093)$ & $0.070(0.045)$ & $0.048(0.033)$ & $0.026(0.019)$\tabularnewline
\hline 
$pp\to W^{-}tjX$ & $11.385$ & $0.418(0.230)$ & $0.215(0.127)$ & $0.144(0.090)$ & $0.071(0.046)$ & $0.046(0.032)$ & $0.025(0.018)$\tabularnewline
\hline 
$pp\to W^{+}bjX$ & $240.342$ & $1.367(0.622)$ & $0.282(0.195)$ & $0.213(0.162)$ & $0.116(0.086)$ & $0.050(0.060)$ & $0.052(0.036)$\tabularnewline
\hline
$pp\to W^{-}\overline{b}jX$ & $180.294$ & $0.970(0.429)$ & $0.227(0.178)$ & $0.225(0.162)$ & $0.181(0.120)$ & $0.088(0.090)$ & $0.052(0.022)$\tabularnewline
\hline
\end{tabular}
\end{table}

For single $b'$ production, the main and most demanding background
is $W^{-}tj$ process. We applied two different sets of criteria to
reduce this background. In the first set, the transverse momentum
cut $p_{T}^{j}>20$ GeV is applied. Applying the transverse momentum
cut ($p_{T}^{j}>50$ GeV) most of the soft jets can be rejected. The
latter cut is more efficient than the first one, since with the former
cut, some soft jets can still be found. Applying the strict cut ($p_{T}^{j}>50$
GeV), reduction efficiency for the background $W^{-}tj$ is high enough
for the interested mass range to observe the $b'$ signal over the
background. 

For the $t'$ signal, we consider $W^{+}Z$, $W^{-}Z$, $W^{+}W^{-}$
and $W^{+}bj$ reducible backgrounds. These are all reducible backgrounds
respect to the applied cuts. The background cross sections with respect
to the applied cuts are displayed in Table \ref{table5}. Here, the
most obstinate background is $W^{+}bj$, which has the cross section
of $240$ pb before the cuts. It decreases to a reasonable value when
we apply the invariant mass and transverse momentum cuts. The background
$W^{-}\overline{b}j$ is also important for $\overline{t}'$ signal.
It should be considered together with the $W^{+}bj$ background, possible
asymmetry for producing single $t'$ and $\overline{t}'$ will give
additional information. By applying the cuts $p_{T}^{j}>50$ GeV and
choosing the appropriate invariant mass intervals $\Delta m=|m_{t'}-m_{Wb}|$
the cross sections for the backgrounds can be reduced below the signal
cross section level.

\section{Signal Significance}

We define the $t'(b')$ signal as the final state including a $b$-jet
or a light jet, a charged lepton (two leptons with opposite sign)
and missing transverse momentum. We also require the reconstructed
invariant mass distribution $m_{WQ}$ showing a peak around $m_{Q'}$
in the interval $300-800$ GeV. We can estimate statistical significance
of the signal by using integrated luminosity of $L_{int}=10^{5}$pb$^{-1}$
that will be achieved in a year at the LHC run. By using integrated
luminosity ($L_{int}$), signal ($\sigma_{S}$) and background ($\sigma_{B}$)
cross sections, the branchings BR($Q'\to WQ$) followed by BR($W\to l\nu$)
to the chosen detectable channel, and the relevant efficiency $\epsilon$
for the signal and background channel, we define statistical significance
(SS) by

\begin{equation}
SS=\frac{\sigma_{S}}{\sqrt{\sigma_{B}}}\sqrt{\epsilon\cdot L_{int}}
\label{eq:8}
\end{equation}
In our calculations, the $b-$tagging efficiency
($60\%$) and identification efficiency for each electron ($90\%$)
and muon ($95\%$) are taken as reference values based on ATLAS studies \cite{atlas99} 
for the relevant final states of signal and background.
Significances are shown in Table \ref{table6} with respect to the
fourth family quark masses in the interval $m_{Q'}=300-800$ GeV for
the three parametrization PI-PIII. In order to calculate signal significances
for all mass ranges (300-800 GeV), one needs to calculate the branching
modes and fractions, and associated total width for each mass values
with proposed CKM options PI, PII and PIII. We use the branchings
from Table \ref{table3} for $b'$ and almost constant values for
$t'$ together with the signal and corresponding background cross
sections in the chosen invariant mass intervals. 

As can be seen from the Table \ref{table6}, in the third option the
statistical significances are very high while the other two options
have fairly low values. The lowest $SS$ values are obtained for the
$t'$ and $\bar{b'}$ productions in the PI option. Both $b'(\bar{b'})$and
$t'(\bar{t'})$ signal can be probed at the LHC at the nominal luminosity
with greater significances for the parametrization PII and PIII. 

As a result, this study shows that if the fourth family quarks are
present, the LHC will discover them and measure their masses. The
single production of fourth family quarks is very important to measure
their mixings with the other families. For the PI-PIII options, $t'$
signal will be observed with a significance greater than $2\sigma$
if $V_{t'q},V_{qb'}\geq0.01$ for interested mass range. As can be
seen from Table \ref{table6}, the other options PII and PIII provide
much better significances for the discovery of fourth family quarks
in the mass interval $300-800$ GeV. 

\begin{table}
\caption{Statistical significance (SS) for $b'(\bar{b'})$ and $t'(\bar{t'})$
at the LHC with the integrated luminosity of $L_{int}=10^{5}$pb$^{-1}.$
\label{table6}}

\begin{tabular}{|c|c|c|c|c|c|c|}
\hline 
SS & \multicolumn{3}{c|}{$b'(\bar{b'})$-quark} & \multicolumn{3}{c|}{$t'(\bar{t'})$-quark}\tabularnewline
\cline{1-1} \cline{5-7} 
\hline 
Mass(GeV) & PI & PII & PIII & PI & PII & PIII\tabularnewline
\hline 
300 & 1.73(0.69) & 43.22(17.27) & 640.91(892.34) & 2.94(1.31) & 73.67(32.86) & 3184.04(2211.22)\tabularnewline
\hline 
400 & 1.78(0.67) & 44.54(16.78) & 544.88(813.12) & 3.34(1.17) & 83.57(29.22) & 3325.72(1928.79)\tabularnewline
\hline 
500 & 1.49(0.51) & 37.45(12.73) & 396.28(597.43) & 2.49(0.75) & 62.26(18.83) & 2290.62(1225.12)\tabularnewline
\hline 
600 & 1.50(0.49) & 37.50(12.17) & 352.11(561.61) & 2.43(0.56) & 60.73(14.13) & 2079.25(909.39)\tabularnewline
\hline 
700 & 1.33(0.39) & 33.19(9.87) & 285.65(453.42) & 2.13(0.44) & 53.48(11.00) & 1707.09(696.92)\tabularnewline
\hline 
800 & 1.34(0.37) & 33.59(9.19) & 261.49(416.81) & 2.07(0.62) & 51.97(15.41) & 1559.32(965.05)\tabularnewline
\hline
\end{tabular}
\end{table}

\section{Conclusions}

This study shows that in case of the single production of fourth family
$b'$ and $t'$ quarks, a mixing down to $1\%$ can be measured at
the LHC with nominal luminosity providing that the fourth family quarks
have mass in the range $300-800$ GeV. The measurements on the $V_{t'q}$
and $V_{qb'}$ can also be improved if a possible luminosity upgrade
is made at the LHC. In case the fourth family quarks are present,
the LHC will discover them in pairs and measure their masses with
a good accuracy. In addition, to obtain the mixing of the fourth family
quarks with the other families, the single production of these new
quarks remains to be measured. Once the LHC begins to run, the Higgs
can be found using the golden mode \cite{arik02} for the relevant
mass range. Given the measurements of cross section for the Higgs
production as well as the branching ratios of Higgs into possible
decay modes, the LHC will be able to verify that the fourth family
quarks do indeed exist. Combining the information from the pair production
of fourth family quarks and their subsequent decays, the single production
mechanism provide a unique measurement of the family mixing with the
fourth family quarks. 

\begin{acknowledgments}
O.C. and H.D.Y acknowledge the support from the Turkish State Planning
Organization under the grant no DPT2006K-120470. H.D.Y's
work is also supported by T\"{U}B\.{I}TAK under the grant no 105T442.
\end{acknowledgments}


\begin{thebibliography}{20}
\bibitem[1]{yao06}W.-M. Yao et al., Journal of Physics G, \textbf{33},
1 (2006);  http://pdglive.lbl.gov/.

\bibitem[2]{cdf05}http://www-cdf.fnal.gov/physics/new/top/2005/ljets/tprime/gen6/public.html.

\bibitem[3]{aaltonen07}T. Aaltonen \emph{et al.} (CDF Collaboration),
Phys. Rev. D \textbf{76}, 072006 (2007), arXiv: hep-ex/0706.3264.

\bibitem[4]{hung08}P.Q. Hung, M. Sher, arXiv:0711.4353.

\bibitem[5]{holdom06}B. Holdom, JHEP \textbf{0608}, 076 (2006); JHEP
\textbf{0703}, 063 (2007); JHEP \textbf{0708}, 069 (2007). 

\bibitem[6]{kribs07}G. D. Kribs et al., Phys. Rev. D \textbf{76},
075016, ANL-HEP-PR-07-39, arXiv: 0706.3718.

\bibitem[7]{celikel95}A. Celikel, A. K. Ciftci, S. Sultansoy, Phys.
Lett. B \textbf{342}, 257 (1995).

\bibitem[8]{atag96}S. Atag et al., Phys. Rev. D \textbf{54}, 5745
(1996).

\bibitem[9]{arik98}E. Arik et al., Phys. Rev. D \textbf{58}, 117701
(1998). 

\bibitem[10]{atlas99}ATLAS Collaboration, ATLAS Detector and Physics
Performance Technical Design Report II, ATLAS-TDR-15, CERN-LHCC-99-15,
Vol 2, p.519, (1999). 

\bibitem[11]{gokhan07}V.E. Ozcan, S. Sultansoy, G. Unel, ATL-COM-PHYS-2007-044.

\bibitem[12]{abazov07}V.M. Abazov \emph{et al.} (D0 Collaboration),
Phys. Rev. Lett. \textbf{98}, 181802 (2007); F. Canelli \emph{et al.}
(CDF Collaboration), Conference Note 8588, 2007.

\bibitem[13]{cteq6m}J. Pumplin et al., CTEQ Collaboration, arXiv:hep-ph/0201195. 

\bibitem[14]{pukhov99}A. Pukhov \emph{et al.}, Preprint INP MSU 98-41/542,
arXiv:hep-ph/9908288; A. Pukhov \emph{et al.}, hep-ph/0412191. 

\bibitem[15]{ocakir07}http://ocakir.web.cern.ch/ocakir/fourthfamily.html

\bibitem[16]{arik02}E. Arik et al., Phys. Rev. D \textbf{66}, 033003
(2002); E. Arik et al., Eur. Phys. J. C \textbf{26}, 9 (2002). 
\end{thebibliography}
\end{document}